# A Dynamic Droplet Breakup Model for Eulerian-Lagrangian Simulation of Liquid-fueled Detonation

Wenhao Wang[a,b], Miao Yang[b], Zongmin Hu[a,*], and Peng Zhang[b,*]

[a]*State Key Laboratory of High-temperature Gas Dynamics (LHD), Institute of Mechanics, Chinese Academy of Sciences, Beijing 100190, China*

[b]*Department of Mechanical Engineering, City University of Hong Kong, Kowloon Tong, Kowloon 999077, Hong Kong*

**Abstract**

This study proposes a dynamic model to reflect the physical image of the droplet breakup process in two-phase detonation flows. This breakup model is implemented in a two-phase detonation solver developed based on an open-source computational fluid dynamic platform, OpenFOAM, and compared with three prevalent models (TAB, PilchErdman, and ReitzKH-RT model) under different droplet diameters (30 – 70 μm) in one- and two-dimensional detonation problems. The simulating results show that the present breakup model well predicts experimentally determined detonation parameters such as detonation velocities and post-wave temperature. In addition, the present model has the advantage of being free of the KH breakup time parameter, which is needed by the ReitzKH-RT model to fit the experimental data. The one-dimensional detonation simulations indicate that different breakup models have a slight impact on the detonation wave velocity because the droplet breakup process does not significantly affect the total heat release as long as it is sufficiently fast to sustain the detonation. However, the two-dimensional detonation simulations show that both the breakup model and the droplet initial diameter significantly affect the detonation cell size due to the different droplet distributions predicted by different models. The breakup length, which is the distance from the shock wave to the location at which sufficiently small child droplets appear, affects the chemical reaction zone thickness, which in turn affects the detonation cell size. A longer breakup length will result in a larger detonation cell size.

---

*Corresponding author.

Peng Zhang: penzhang@cityu.edu.hk

Zongmin Hu: huzm@imech.ac.cn

**Nomenclature**

| | |
|---|---|
| $A_p$ | surface area of droplet (m$^2$) |
| $B_1$ | KH breakup time coefficient (-) |
| $B_M$ | mass transfer number (-) |
| $D_0$ | initial droplet diameter (m) |
| $D$ | parent droplet diameter (m) |
| $D_f$ | diffusion coefficient (m$^2$/s) |
| $D_s$ | child droplet's diameter (m) |
| $D_{32}$ | Sauter mean diameter (m) |
| $E$ | total energy (J) |
| $e$ | internal energy (J) |
| $n(t)$ | droplet number (-) |
| $t_b$ | breakup completion time (s) |
| $t_i$ | initial breakup time (s) |

*Greek letter*

| | |
|---|---|
| $\alpha$ | coefficient in the present model (-) |
| $\rho$ | Density (kg/m$^3$) |
| $\mu$ | dynamic viscosity (N·s/m$^2$) |
| $\nu$ | kinematic viscosity coefficient (m$^2$/s) |
| $\psi$ | convective term |

**Abbreviation**

| | |
|---|---|
| $c_p$ | heat capacity of droplet (J/K) |
| $c_g$ | heat capacity of gas phase (J/K) |
| $C_d$ | drag coefficient (-) |
| $U_D$ | detonation velocity (m/s) |
| $d_p$ | droplet diameter (m) |
| $h_c$ | convective heat transfer coefficient (W/m$^2$) |
| $k_{fk}$ | forward reaction rate constant (-) |
| $k_{bk}$ | reverse reaction rate constant (-) |
| $MW_i$ | molecular weight of species $i$ (kg/mol) |
| $m_p$ | mass of droplet (kg) |
| $nq$ | number of chemical reactions (-) |
| $ns$ | number of species (-) |
| $N_p$ | particle number in a cell (-) |
| $Nu$ | Nusselt number: $Nu = hL/k$ |
| $Oh$ | Ohnesorge number: $Oh = \sqrt{We_p}/Re_p$ |
| $Pr$ | Prandtl number: $Pr = c_p\mu/k$ |
| $p$ | pressure (Pa) |
| $\dot{Q}_c$ | convective heat transport rate (J/s) |
| $\dot{Q}_{latent}$ | latent heat transport rate (J/s) |
| $R$ | specific gas constant (J/kg/K) |
| $R_u$ | universal gas constant (J/mol/K) |
| $Re_p$ | droplet Reynold number: $Re_p = \rho_p u_r D/2\mu$ |
| $Sh$ | Sherwood number: $Sh = hL/D_f$ |
| $Sc$ | Schmidt number: $Sc = \nu/\overline{D}_f$ |
| $\boldsymbol{S_p}$ | source term from liquid phase |
| $T$ | gas phase temperature (K) |
| $T_a$ | Talyor number: $T_a = Oh/\sqrt{We_p}$ |
| $T_p$ | droplet temperature (K) |
| $U_{CJ}$ | Chapman and Jouguet detonation velocity (m/s) |
| $U_D$ | detonation velocity (m/s) |
| $\boldsymbol{u}$ | gas phase velocity vector (m/s) |
| $u_r$ | magnitude of relative velocity (m/s) |
| $\boldsymbol{u_p}$ | droplet velocity vector (m/s) |
| $V_c$ | calculation cell volume (m$^3$) |
| $We$ | Weber number: $We = \rho u_r D_p/2\sigma_p$ |
| $We_p$ | Weber number of droplet: $We_p = \rho_l u_r D_p/2\sigma_p$ |
| $We_c$ | critical Weber number: $We_c = \rho_l u_r D_s/2\sigma_p$ |
| $Y_i$ | mass fraction of species $i$ (-) |
| $Y_s$ | mass fraction of fuel vapor at the droplets' surface (-) |
| $Y_g$ | mass fraction of fuel vapor at the ambient gas (-) |
| $\rho_p$ | droplet density (kg/m$^3$) |
| $\Lambda_{RT}$ | Rayleigh–Taylor wavelength (m) |
| $\Omega_{RT}$ | Rayleigh–Taylor wave frequency (s$^{-1}$) |
| $\Lambda_{KH}$ | Kelvin–Helmholtz wavelength (m) |
| $\Omega_{KH}$ | Kelvin–Helmholtz wave frequency (s$^{-1}$) |
| $\sigma_p$ | droplet surface tension (N/m) |
| $\tau_{RT}$ | RT breakup characteristic time (s) |
| $\tau_{KH}$ | KH breakup characteristic time (s) |
| $\nu'_{ik}$ | stoichiometries of species $i$ for forward reaction (-) |
| $\nu''_{ik}$ | stoichiometries of species $i$ for backward reaction (-) |
| $\lambda_c$ | detonation cell size (m) |

# 1. Introduction

Detonation is a combustion wave phenomenon with a moving shock sustained by a followed chemical reaction zone. While detonation poses risks in both industrial and everyday settings due to the explosion of combustible gases and dust, it also holds potential for application in propulsion systems due to its exceptional thermodynamic cycle performance [1-3]. Conventional detonation engines, such as pulse, rotating, and oblique detonation engines, commonly employ liquid hydrocarbon fuels as propellants due to their convenient storage and high energy density. However, compared with the extensively studied gas-fueled detonations, numerical simulations of liquid-fueled detonations face additional challenges in dealing with various intricate gas-liquid interaction processes, such as droplet breakup, collision, and evaporation.

In most two-phase flow simulations, the gas phase is treated as a continuum, as referred to be an Eulerian approach. Two prevalent approaches are raised to deal with the liquid phase in two-phase flow simulation: the Eulerian and Lagrangian approaches. The Eulerian approach treats the liquid phase as a continuum. It is less computationally demanding but requires models to describe the discrete droplets. Hayashi et al. [4] used this approach to simulate JP-10/air two-phase detonation in a rotating detonation engine, but they ignored the droplet size distributions, droplet breakup, and droplet collisions in the simulation. Jourdaine et al. [5] assumed the liquid spray to be comprised of droplets defined by specific properties and used a number density function to describe the dispersed liquid phase, where the PilchErdman breakup model was adopted. The number of droplet diameter samples to consider the size polydispersity of the droplets may significantly influence the detonation wave structure in their method. The Lagrangian approach treats droplets as discrete phases, tracks droplets' motion, and models droplet breakup and collision. Ren et al. [6] used a Lagrangian approach to simulate the stationary problem of two-phase oblique detonation waves, but they ignored the breakup of droplets. Zhang et al. [7] developed a hybrid Eulerian-Lagrangian solver named RYrhoCentralFoam based on the OpenFOAM for two-phase detonation, in which no breakup models were used because of the small droplet diameter. In the subsequent studies [8, 9], several models, particularly the PilchErdman model were adopted. Musick et al. [10] investigated the droplet evaporation and breakup effects on heterogeneous detonations, considering the difference between WERT49 and ReitzKH-RT breakup models.

From the above discussions, it can be seen that existing studies on liquid-fueled detonations have either neglected droplet breakup or adopted prevalent droplet breakup models that were established not for detonations or supersonic combustion. Therefore, the present research was motivated by the urgent need to develop physically correct droplet breakup models for two-phase detonations and supersonic combustion. Modeling droplet breakup in two-phase flow simulation is conventionally based on parametrizing the droplet fragmentation processes observed in experiments. Guildenbecher et al. [11] classified the breakup of a Newtonian droplet into five regimes according to the different Weber numbers, determined from the earlier experiments by Hsiang et al. [12] and Pilch et al. [13], as shown in Fig.1. In this classification, a droplet will experience a 'catastrophic' breakup process when the Weber number is higher than 350. However, with advances in experimental techniques, the catastrophic breakup was found to be an artifact of shadowgraphs [14-17]. Theofanous et al. [16] use laser-induced fluorescence to visualize liquid droplet breakup after a shock wave and found that some experiments based on the shadowgraph method may lead to misinterpretations about the breakup mechanism, the 'catastrophic' breakup was a mirage of the shadowgraphs used to "visualize" waves.

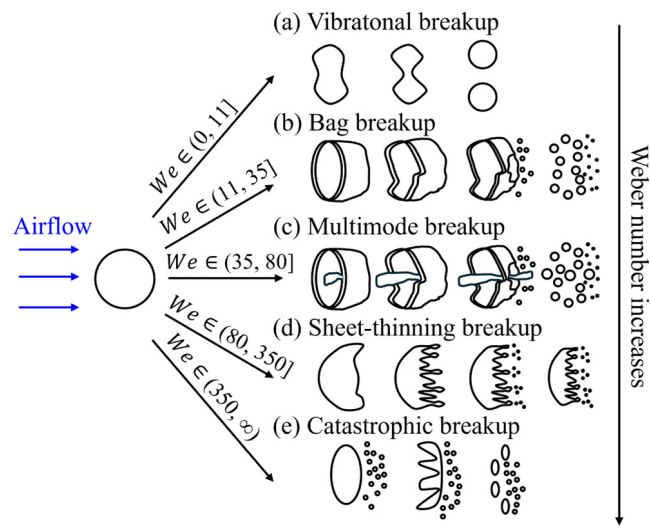

Fig. 1 Nomogram of droplet breakup regimes classified by the Weber number [11]

In most previous models of droplet breakup, the three parameters are often used: the initial breakup time $t_i$, the breakup completion time $t_b$, and the child droplet size $D_s$ [18]. The initial breakup time is the duration from the droplet being exposed to the gas flow to the moment when the droplet begins to break, during which it experiences the deformation process. The breakup completion time represents the total time of the breakup process from the droplet being exposed to the gas flow to the moment when all the child droplets stop breaking. The distribution of child droplets may not be monodispersed, while many models often assume that the child droplets have the same diameter for simplification.

In general, the existing droplet breakup models may fall into two approaches: those that utilize physical analogies to construct the breakup process and those that rely on experiments to model the breakup process. The first approach originates in the TAB (Taylor Analogy Breakup) model suggested by Talyor [19] and further developed by O'Rourke et al. [20]. The TAB model uses an analogy between the oscillation and distortion of a droplet and a spring-mass system to build a second-order ordinary differential equation of the displacement of the equator of the droplet from its equilibrium position, and the parameters in the equation are based on experimental or theoretical results. In The TAB model, the droplet instantly breaks up into several child droplets when it deforms to a certain extent, namely $t_b = t_i$. In addition, the child droplets' diameter $D_s$ is determined by energy conservation, so the number of child droplets is a constant which can be obtained from mass conservation. Tanner et al. [21] extended the TAB model to the ETAB

(Extended Taylor Analogy Breakup) model and considered the breakup process as an exponential decline of droplets' diameter. In the ETAB model, $t_i$, $t_b$, and $D_s$ are the same as those in the TAB model, but the rate of the droplet number, $\frac{dn(t)}{dt} = 3K_{br}n(t)$, where the constant $K_{br}$ depends on the breakup regimes.

The second approach is mainly based on the experimental delineation of different droplet breakup regimes. In these models, the droplets' diameter is considered to decrease continuously as

$$\frac{dD}{dt} = \frac{D - D_s}{\tau_b} \qquad (1.1)$$

in which the characteristic time $\tau_b$ is modeled depending on different breakup regimes, and $D_s = \frac{2We_c\sigma_p}{\rho u_r}$ is calculated by the experimentally-determined critical breakup Weber number, namely. The major difference among these experimental models is the way to model $\tau_b$, which is a segmented function with respect to the Weber number. Reitz-Diwakar model considered two breakup regimes, bag and stripping breakup [22, 23]. PilchErdman model considers vibrational, bag, bag-and-stamen, sheet, wave crest, and catastrophic breakup regimes [13]. Maier and Witting model (referred to as SHF in the literature) considers bag, multimode, and shear breakup regimes, while this model assumes that the child droplets' diameter follows normal distribution [24]. Patterson and Reitz raised a hybrid KH-RT model based on the two instability waves, the KHI (Kelvin–Helmholtz instability) wave and the RTI (Rayleigh–Taylor instability) wave [25]. This model treats the breakup process as a competition between the KHI and the RTI, where the RT breakup process is instantaneous, but the KH breakup is modeled as a process in which the diameter of the parent droplet decreases gradually following Equ. (1.1) and tiny droplets are exfoliated stepwise.

This work concerns the two-phase flow simulation of liquid-fueled denotation, in which droplets break up due to the strong shock wave and the following high-speed flow. Theofanous et al. [16] summarized and divided the breakup process into the Rayleigh–Taylor piercing (RTP) and Shear-Induced (SHI) breakup caused by the Kelvin–Helmholtz waves. Sharma et al. [15] used shadowgraph and pulsed laser to investigate the shock–droplet interaction, which was divided into the initial wave stage and the droplet breakup stage. The initial wave stage characterizes the formation of various wave structures (reflected waves, transmitted waves, and refracted waves), and the droplet breakup stage characterizes the dynamic process of droplet breakup. In the breakup stage, they also found two breakup modes, namely the RTP and KHI modes. In the RTP mode, the parent droplet undergoes complete deformation into a flattened disc, on which the RT instability wave forms and leads to child droplet formation quickly. In the KHI mode, the KH instability wave originates from the windward surface of the droplet, forming a thin sheet that undergoes rupture and results in a dynamic process of gradual daughter droplets/mist generation.

From the above review of the existing models and experiments, it can be concluded that the previous models do not completely reflect the physical processes of droplet breakup after shock waves or within supersonic flows. In addition, the effects of different breakup models on two-phase detonation simulation needs further clarifications. To address these issues, this work establishes an open-source two-phase compressible flow solver based on OpenFOAM and extends the ReitzKH-RT model to higher Weber numbers based on the physical considerations of droplet breakup in supersonic flows. Moreover, we choose three existing representative breakup models together with the dynamical breakup model presented in this work to conduct a comparative study on the effects of different breakup models on the two-phase detonations. The following text will be organized as follows. The computational methodology and modeling will be presented in Section 2, followed by the presentation of the present dynamic model and computational results in Section 3.

## 2. Computational Methodology and Modeling

### 2.1. Governing Equations and Models of Eulerian Phase

This study employs the Euler approach to simulate the gas flow. The governing equation for the compressible flow can be written as

$$\frac{\partial \mathbf{U}}{\partial t} + \frac{\partial (\mathbf{E} - \mathbf{E}_v)}{\partial x} + \frac{\partial (\mathbf{F} - \mathbf{F}_v)}{\partial y} + \frac{\partial (\mathbf{G} - \mathbf{G}_v)}{\partial z} = \mathbf{S} + \mathbf{S}_p, \qquad (2.1)$$

in which $\mathbf{U}$ is the conserved quantity containing the mass, momentum, energy density, and gas mixture components of the control volume,

$$U = [\rho, \rho u, \rho v, \rho w, \rho E, \rho Y_1, \dots, \rho Y_{ns-1}]^T, \tag{2.2}$$

$E, F$, and $G$ denote the momentum flux per unit area of the surface of the control body along the $x$, $y$, and $z$ directions,

$$\begin{cases} E = [\rho u, \rho u^2 + p, \rho uv, \rho uw, (E+p)u, \rho u Y_1, \dots, \rho u Y_{ns-1}]^T \\ F = [\rho v, \rho uv, \rho v^2 + p, \rho vw, (E+p)v, \rho v Y_1, \dots, \rho v Y_{ns-1}]^T \\ G = [\rho w, \rho uw, \rho vw, \rho w^2 + p, (E+p)w, \rho w Y_1, \dots, \rho w Y_{ns-1}]^T \end{cases}, \tag{2.3}$$

$E_v, F_v, G_v$ denote the viscous stress,

$$\begin{cases} E_v = \left[0, \tau_{xx}, \tau_{xy}, \tau_{xz}, u_i\tau_{xi} + q_x, \rho D_f \frac{\partial Y_1}{\partial x}, \dots, \rho D_f \frac{\partial Y_{ns-1}}{\partial x}\right]^T \\ F_v = \left[0, \tau_{yx}, \tau_{yy}, \tau_{yz}, u_i\tau_{yi} + q_y, \rho D_f \frac{\partial Y_1}{\partial y}, \dots, \rho D_f \frac{\partial Y_{ns-1}}{\partial y}\right]^T \\ G_v = \left[0, \tau_{zx}, \tau_{zy}, \tau_{zz}, u_i\tau_{zi} + q_z, \rho D_f \frac{\partial Y_1}{\partial z}, \dots, \rho D_f \frac{\partial Y_{ns-1}}{\partial z}\right]^T \end{cases}, \tag{2.4}$$

$S$ and $S_p$ are the source terms from chemical reactions and two-phase interactions,

$$\begin{cases} S = [0,0,0,0,0,\dot{\omega}_1, \dots, \dot{\omega}_{ns-1}]^T \\ S_p = [S_{mass}, S_{mom}, S_{energy}, S_{species}]^T \end{cases}. \tag{2.5}$$

In Equ. (2.1) - (2.5), $u, v, w$ represent the gas velocity in each direction, $ns$ is the number of species, and $Y_1, \dots, Y_{ns-1}$ are the mass fractions of each species. $\rho$ and $p$ are the density and pressure of gas, satisfying the ideal gas law $p = \rho RT$. $R$ is the specific gas constant calculated by $R = R_u \sum_{i=1}^{ns} \frac{Y_i}{MW_i}$, where $R_u$ is the universal gas constant, and $MW_i$ is the molecular weight of the $i$-th species. $E = e + \frac{1}{2}(u^2 + v^2 + w^2)$ is the total energy, in which $e$ is the internal energy. $\tau_{ij}$ $(i,j = 1,2,3)$ is the stress expressed as $\tau_{ij} = -\frac{2}{3}\mu\delta_{ij}u_{k,k} + \mu(u_{i,j} + u_{j,i})$, and $\mu$ is the dynamic viscosity calculated by the Sutherland's law. $q$ is the diffusive heat flux. From Fourier's law, $q = -k\nabla T$ with $k$ being the thermal conductivity. $D_f$ is the diffusion coefficient.

In the source term $\boldsymbol{S}$, $\dot{\omega}_i$ is the net production of $i$-th species. For a chemical reaction with $ns$ chemical components and $nq$ radical reaction equations,

$$\dot{\omega}_i = MW_i \sum_{k=1}^{nq} (v''_{ik} - v'_{ik}) \left[\sum_{i=1}^{ns} (\alpha_{ik} c_{\chi i})\right] \cdot \left[k_{fk} \prod_{i=1}^{ns} (c_{\chi i})^{v'_{ik}} - k_{bk} \prod_{i=1}^{ns} (c_{\chi i})^{v''_{ik}}\right]. \tag{2.6}$$

In (2.6), $v'_{ik}$ and $v''_{ik}$ are the stoichiometries of the $i$-th species before and after the reaction in the $k$-th reaction, respectively. $\alpha_{ik}$ is the three-body effect coefficient of $i$-th species in $k$-th reaction, $c_{\chi i}$ is molar concentration of $i$-th species. $k_{fk}$ and $k_{bk}$ are forward and backward reaction rate constants, respectively. $\boldsymbol{S_p}$ will be explained in the following section.

## 2.2. Governing Equations and Models of Lagrangian Phase

This study utilizes the Lagrangian particle tracking (LPT) method to track the motion of liquid droplets, which is described by the following equations,

$$\frac{dm_p}{dt} = \dot{m}_p, \tag{2.7}$$

$$\frac{d\boldsymbol{u}_p}{dt} = \frac{\boldsymbol{F}_p}{m_p}, \tag{2.8}$$

$$\text{and } c_p \frac{dT_p}{dt} = \frac{\dot{Q}_c + \dot{Q}_{latent}}{m_p}. \tag{2.9}$$

where $m_p$, $\boldsymbol{u}_p$, and $T_p$ are the mass, velocity, and temperature of each droplet, $c_p$ is the heat capacity of a droplet with constant pressure. The droplet is treated as a sphere of the mass $m_p = \frac{1}{6}\pi\rho_p D_p^3$, where $\rho_p$ and $D_p$ are the density and diameter of the droplet.

In the two-phase supersonic flow, the convective transport is considered by using the so-called "film theory" [26]. This theory assumes that the resistance of mass exchange between the moving droplets and gas can be modeled by introducing the concept of gas films of constant thickness, and gives the expression $\dot{m}_p = -\pi D_p Sh \overline{D}_f \bar{\rho} \ln(1 + B_M)$, where $\bar{\rho}$ and $\overline{D}_f$ are the average density and binary diffusion coefficient of the gas mixture in the films. $Sh$ is Sherwood number modelled as $Sh = 2.0 + 0.6 Re_p^{\frac{1}{2}} Sc^{\frac{1}{3}}$. $Re_p \equiv \frac{\rho_p D_p u_r}{\mu}$ is the Reynolds number of the droplet, where $u_r$ is the magnitude of relative velocity of gas and droplet. $Sc = \frac{\nu}{D_f}$ is the Schmidt number. $\nu = \frac{\mu}{\rho}$ is the kinematic viscosity coefficient, $B_M = \frac{Y_s - Y_g}{1 - Y_s}$ is the mass transfer number, $Y_s$ and $Y_g$ are the mass fraction of fuel vapor at the droplets' surface and ambient gas.

$\boldsymbol{F}_p$ is calculated by sphere drag model [27] and expressed as

$$\boldsymbol{F}_p = -\frac{18\mu}{\rho_p D_p^2} \frac{C_d Re_p}{24} m_p (\boldsymbol{u}_p - \boldsymbol{u}), \qquad (2.10)$$

where the drag coefficient $C_d$ is modelled as

$$C_d = \begin{cases} 24\left(1 + \frac{1}{6} Re_p^{\frac{2}{3}}\right), & Re_p \leq 1000 \\ 0.424, & Re_p > 1000 \end{cases}. \qquad (2.11)$$

The convective heat transfer is calculated by

$$\dot{Q}_c = h_c A_p (T - T_p), \qquad (2.12)$$

where $A_p = \pi D_p^2$ is the surface area of the droplet. $h_c \equiv \frac{Nu \cdot k}{D_p}$ is the convective heat transfer coefficient, in which $Nu$ is the Nusselt number calculated as

$Nu = 2.0 + 0.6Re_p^{\frac{1}{2}}Pr^{\frac{1}{3}}$ by Ranz-Marshall model[28]. $Pr = \frac{c_p\mu}{k}$ is the Prandtl number, $c_p$ is the heat capacity with constant pressure of gas phase. $\dot{Q}_{latent}$ is the latent heat of evaporation,

$$\dot{Q}_{latent} = -\dot{m}_p h_l(T_p), \qquad (2.13)$$

where $h_l(T_p)$ is the latent heat of droplet at $T_p$. Using the above liquid parameters, we can obtain each term of $\boldsymbol{S}_p$ in each calculation cell with volume $V_c$ and $N_p$ particles in it,

$$S_{mass} = -\frac{1}{V_c}\sum_1^{N_p} \dot{m}_p, \qquad (2.14)$$

$$\boldsymbol{S}_{mom} = -\frac{1}{V_c}\sum_1^{N_p} \boldsymbol{F}_p, \qquad (2.15)$$

$$S_{energy} = -\frac{1}{V_c}\sum_1^{N_p} (\dot{Q}_c + \dot{Q}_{latent}), \qquad (2.16)$$

$$\text{and } S_{species,i} = \begin{cases} -\frac{1}{V_c}\sum_1^{N_p} \dot{m}_p, & condensed\ species \\ 0, & other\ species \end{cases}. \qquad (2.17)$$

## 2.3. Existing Models of Droplet Breakup

Three existing models of droplet breakup (TAB, PilchErdman, and ReitzKH-RT) are considered in this work and shown in Fig.2.

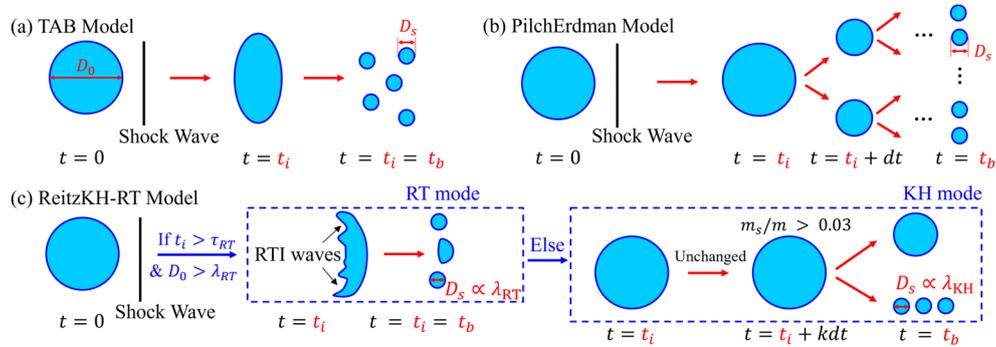

Fig. 2 The underlying physical processes and parameter relationships of the (a) TAB, (b) PilchErdman, and (c) ReitzKH-RT models.

*2.3.1 TAB model*

In the TAB model, the droplets will undergo two main processes: a deformation process for a period of time and an instantaneous breakup [20]. Taking $x$ to be the displacement of the equator of the droplet from its equilibrium position and using analogy of the droplet distortion to a spring-mass system, we have

$$m_p \ddot{x} = F - kx - d\dot{x}, \qquad (2.18)$$

in which

$$\begin{cases} \dfrac{F}{m_p} = 2C_F \dfrac{\rho_g u^2}{\rho_p D} \\ \dfrac{k}{m_p} = 2C_k \dfrac{\sigma}{\rho_p D^3} \\ \dfrac{d}{m_p} = 2C_d \dfrac{\mu_l}{\rho_p D^3} \end{cases}. \qquad (2.19)$$

$C_F, C_k$, and $C_d$ are determined by the previous experimental and theoretical results. The breakup happens when $x > \frac{1}{2} C_b D$, in which $C_b$ is an artificially given parameter, and $D$ is the diameter of the parent droplet. Usually, $C_k = 8$, $C_d = 5$, $C_F = 1/3$, $C_b = 0.5$. Let $y = \frac{2x}{C_b D}$ to be the nondimensional displacement and according to shock experiments, Equ. (2.18) can be solved as $y(t) = \frac{C_F}{C_k C_b} We(1 - \cos \omega t)$, where $\omega = 8C_k \frac{\sigma_p}{\rho_p D^3}$ and $We = \frac{\rho_g u_r D}{2\sigma_p}$. When $t = t_b$, the parent droplets suddenly break into several child droplets, as shown in Fig.2 (a), where $t_b$ can be calculated by $y(t_b) = 1$. The child droplets' diameter $D_s$ is calculated based on the energy conservation as $\frac{D_s}{D_{32}} = 1 + \frac{8K}{20} + \frac{\rho_p r^3}{\sigma} \dot{y} \left( \frac{6K-5}{120} \right)$, in which $K$ is determined by experiments, and $D_{32}$ is determined by $\frac{We}{\sqrt{Re_p}} = \frac{\rho u_r^2 D_{32}}{\sigma_p}\sqrt{\frac{2\nu}{u_r D_{32}}} = 1$.

*2.3.2 PilchErdman model*

In the PilchErdman model, the breakup occurs only when $We > We_c$, where $We_c = 12(1.0 + 1.077 Oh^{1.6})$. When breakup begins, the parent droplets will experience a "chain breakup" process as shown in Fig.2 (b), which means that the diameter of each child droplet in a droplet parcel will decrease with each time step as

$$\frac{dD}{dt} = -\frac{D - D_s}{\tau_b}, \qquad (2.20)$$

and the number of droplets $n$ will increase with each time step and is determined by the mass conservation as

$$n(t)D^3(t) = n(t + dt)D^3(t + dt). \qquad (2.21)$$

In (2.20), $\tau_b = \tau_{b0}\frac{D}{u_r}\sqrt{\frac{\rho_p}{\rho}}$ and

$$\tau_{b0} = \begin{cases} 6.0(We - 12.0)^{-0.25}, & 12 < We \leq 18 \\ 2.45(We - 12.0)^{-0.25}, & 18 < We \leq 45 \\ 14.1(We - 12.0)^{-0.25}, & 45 < We \leq 351 \\ 0.766(We - 12.0)^{-0.25}, & 351 < We \leq 2670 \\ 5.5, & We \geq 2670 \end{cases} \qquad (2.22)$$

$D_s$ is the diameter of stable child droplets given by $D_s = We_c \frac{\sigma_p}{\rho u_r^2}\left(1 - \frac{V_d}{u_r}\right)^{-2}$, where $V_d = |\boldsymbol{u}_p - \boldsymbol{u}|\sqrt{\frac{\rho_g}{\rho_p}}(C_1\tau_{b0} + C_2\sqrt{\tau_{b0}})$. In compressible flows, $C_1 = 0.375$ and $C_2 = 0.2274$.

*2.3.3 ReitzKH-RT model*

The ReitzKH-RT model considers the breakup process as the competition between the KH instability and RT instability waves [29]. According to the linear stability analysis [30], the wavelength and frequency of the RT wave is

$$\Lambda_{RT} = 2\pi \sqrt{\frac{3\sigma_p}{|g_t(\rho - \rho_p)|}}, \quad (2.23)$$

$$\Omega_{RT} = \sqrt{\frac{2}{3\sqrt{3}\sigma_p} \frac{[-g_t(\rho - \rho_p)]^{1.5}}{\rho + \rho_p}}, \quad (2.24)$$

where $g_t$ is the magnitude of acceleration in the direction of droplets' motion. In the RT breakup process, the parent droplets are considered to undergo a gradual process of deformation until $\tau_{RT}$, expressed as $\tau_{RT} = \frac{C_\tau}{\Omega_{RT}}$, in which $C_\tau$ is an adjustable parameter. Then the parent droplet suddenly breaks into child droplets with diameter with a diameter of RT wavelength level as shown in Fig.2 (c). Similarly, in the KH breakup process, the wavelength and frequency of the KHI wave is

$$\Lambda_{KH} = 9.02D \frac{(1 + 0.45Oh^{0.5})(1 + 0.4T_a^{0.7})}{2(1 + 0.87We^{1.67})^{0.6}}, \quad (2.25)$$

$$\Omega_{KH} = \frac{(0.34 + 0.38We^{1.5})}{(1 + Oh)(1 + 1.4T_a^{0.6})} \cdot \left[\frac{\rho_p D^3}{8\sigma_p}\right]^{-0.5}, \quad (2.26)$$

where the Ohnesorge number is defined as $Oh = \frac{\sqrt{We_p}}{Re_p}$, the Taylor number is $T_a = \frac{Oh}{\sqrt{We_p}}$, the liquid Weber number is $We_p = \frac{\rho_l u_r^2 D}{2\sigma_p}$. In the KH breakup process, the diameter of the parent droplet $D$ satisfies

$$\frac{dD}{dt} = -\frac{D - D_s}{\tau_{KH}}, \quad (2.27)$$

where $\tau_{KH} = \frac{3.726 B_1 D}{2\Omega_{KH}\Lambda_{KH}}$ and the diameter of the child droplets $D_s = 2B_0\Lambda_{KH}$. The reduced mass will form the child droplets when it accumulates to a certain value (usually 0.03 of the parent droplet's mass) as Fig.2 (c), which describes a stepwise process.

## 2.4. Computational Specifications

### 2.4.1 Open-source solver based on OpenFOAM

We developed an open-source solver named detonationSprayFoam based on the rhoCentralFoam in OpenFOAM V7 [31] by introducing the LPT method and chemical combustion for the two-phase detonation concerned in this work. A similar flow solver RYrhoCentralFoam has been developed by Zhang et al.[7] for two-phase supersonic combustion and detonation. Unfortunately, their solver is not currently available as an open-source one. Consequently, a direct comparison of the two solvers is impossible in the present work. In our solver, the finite volume method (FVM) is used to solve the governing equations of gas phase. The first-order Euler scheme is used for the temporal discretization and the "Gauss-Limited linear" schemes are used for spatial derivatives. The KNP scheme is utilized to capture the shock wave [32]. Use $\boldsymbol{\psi}$ to represent the convective terms such as $\rho\boldsymbol{u}$, we can integrate it in a control volume as

$$\int_V \nabla \cdot [\boldsymbol{u}\boldsymbol{\psi}]\, dV = \int_S d\boldsymbol{S} \cdot [\boldsymbol{u}\boldsymbol{\psi}] \approx \sum_f \boldsymbol{S}_f \cdot \boldsymbol{u}_f \boldsymbol{\psi}_f = \sum_f \phi_f \boldsymbol{\psi}_f, \quad (2.28)$$

in which index $f$ represents each surface of the control volume. To calculate the integration of the convective term, the flux is split into two directions corresponding to the flow outward or inward of the face as

$$\sum_f \phi_f \psi_f = \sum_f [\alpha \phi_{f+}\psi_{f+} + (1 - \alpha_0)\phi_{f-}\psi_{f-} + \omega_f(\psi_{f-} - \psi_{f+})]. \quad (2.29)$$

The volume flux can be calculated as $\psi_{f\pm} = \max(c_{f+}|S_f| \pm \phi_{f+}, c_{f-}|S_f| \pm \phi_{f-}, 0)$, and $c_{f\pm} = \sqrt{\gamma R T_{f\pm}}$ are the speed of gas at the cell face, outward or

inward of the owner cell. For the KNP method, $\alpha_0 = \frac{\psi_{f+}}{\psi_{f+}+\psi_{f-}}$, the weighting coefficient $\omega_f = \alpha_0(1-\alpha_0)(\psi_{f+}+\psi_{f-})$.

*2.4.2 Cases Setting*

In this work, we consider the two-phase detonation of n-heptane with different droplet diameters ($D_0 = 30 - 70$ μm) and different droplet breakup models. Notice that, in OpenFOAM, the computational domain is always three-dimensional, "empty" boundary conditions are used for one- and two-dimensional cases. Fig.3 (a) shows the computational domain of one-dimensional calculations. The length of the domain is 1 m, with 2×10⁵ monodisperse droplet parcels, each of which has different particle numbers in a parcel depending on the droplets' diameter, ensuring that the total mass of n-heptane and air satisfies the fuel/oxidizer ratio under stoichiometry condition. The total number of meshes is 1×10⁴ with the mesh size $\Delta x = 0.10$ mm. An ignition region of 2 mm in length with an equivalent ratio of gaseous n-heptane to air is used to ignite the detonation wave. The left boundary is set as the wall, the right boundary is set as the supersonic outflow with the zero-gradient condition for each variable, and the other boundaries are set as empty. The CFL number is set as 0.1 for gas-phase flow.

Fig.3 (b) shows the computational domain of two-dimensional calculations. An ignition area with a 5 mm length is set with an equivalent ratio of gaseous n-heptane to air and three high-temperature and high-pressure regimes. The length of the computational domain is 0.5 m, the width is 0.53 m, and the grid size is 0.15 mm in both directions, generating a mesh of $1.18\times10^7$ grid cells. $1\times10^6$ monodisperse n-heptane droplet parcels are uniformly distributed in the computational domain, and the total mass of n-heptane and air also satisfies the fuel/oxidizer ratio under the stoichiometry condition. The left and right boundaries are the same as in the one-dimensional case, the up and down boundaries are set as slip walls, and the front and back boundaries are empty. For chemical reactions, a skeletal mechanism of n-heptane combustion with 46 species and 115 reactions [33] has been validated in many previous works and was used in the present work.

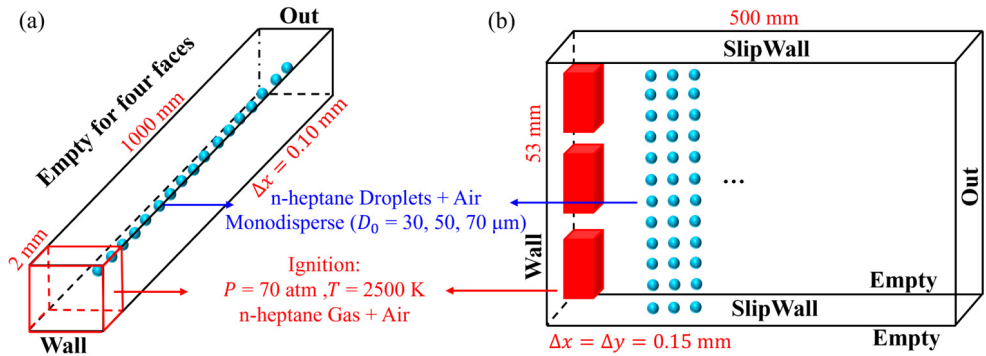

Fig. 3 The computational domains for the (a) one-dimensional and (b) two-dimensional n-heptane two-phase detonation.

### 2.4.3 Mesh-independence Verification

To demonstrate the mesh independence of the present two-phase detonation simulations, a series simulations were conducted by setting different mesh sizes. For one-dimensional simulations, three mesh sizes (0.15, 0.10, and 0.05 mm) were used with the PilchErdman breakup model. Table 1 gives the detonation velocity for different meshes. All these meshes give similar velocities with the difference being less than 1%, implying that these results all show good grid independence. In the following one-dimensional simulations, we choose the mesh with 0.10 mm grid size.

For two-dimensional simulations, three mesh sizes (0.20, 0.15, and 0.10 mm) were used. Table 2 gives the detonation propagation velocity and the cell size of these cases. The results show that the detonation wave cell sizes $\lambda_c$ and detonation velocity $U_D$ are similar at grid sizes of 0.15 and 0.10 mm, while the results with a grid size of 0.2 mm are slightly larger than them by 2%. Therefore, it can be concluded that a grid size of 0.15 mm is sufficient to calculate the two-phase detonation problem, and the following two-dimensional simulations are set with this grid size. Furthermore, it is seen that the detonation velocities obtained in the one- and two-dimensional simulations are very close with the discrepancy being about 0.3%.

|       | $\Delta x = 0.15$ mm | $\Delta x = 0.10$ mm | $\Delta x = 0.05$ mm |
|-------|----------------------|----------------------|----------------------|
| $U_D$ | 1798 m/s             | 1805 m/s             | 1809 m/s             |

Table 1 The detonation propagation velocity under different mesh sizes of one-dimensional simulations

|             | $\Delta x = 0.20$ mm | $\Delta x = 0.15$ mm | $\Delta x = 0.10$ mm |
|-------------|----------------------|----------------------|----------------------|
| $U_D$       | 1822 m/s             | 1799 m/s             | 1785 m/s             |
| $\lambda_c$ | 35.3 mm              | 30.6 mm              | 30.9 mm              |

Table 2 The detonation propagation velocity and detonation cell sizes for different mesh sizes.

## 3. Results and Discussion

### 3.1. *Dynamical Model for droplet breakup in supersonic flows*

Recent experiments have shown that droplet fragmentation is dominated by the Kelvin-Helmholtz instability (KHI) at high $We$s [15]. This process involves the formation of KH waves on the droplet surface, which gradually moves towards the equator and peels off child droplets through sheet fragmentation. To describe this gradual stripping of child droplets over time, we developed a dynamic model consisting of a system of ordinary differential equations (ODEs) without the need for characteristic time as an adjustable parameter, extending the ReitzKH-RT model to the higher $We$s.

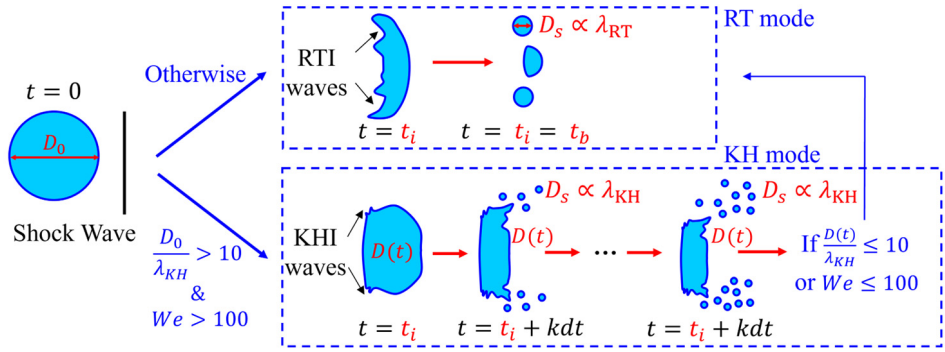

Fig. 4 The underlying physical processes of the present droplet breakup model

In the present model, the KH breakup occurs when $We > 100$ and $\frac{D_0}{\lambda_{KH}} > 10$, otherwise the RT breakup happens, based on the experimental observations [15]. The RT breakup process is the same as that in the ReitzKH-RT model, implying our model degenerates to the ReitzKH-RT model for low $We$ numbers. When it comes to the KH breakup process, $t_i$ is assumed to be 0, child droplets are continuously exfoliated over time [15], as shown in Fig.4 for the schematic of this process.

According to the mass conservation of droplets in a parcel, we have

$$N \cdot D_0^3 = N \cdot D^3(t) + n(t) \cdot D_s^3(t), \qquad (3.1)$$

where $D_0$ is the initial diameter of parent droplets, $N$ represents the number of

parent droplets in a droplet parcel which remains constant in KH breakup process. From Equ. (2.25), we have

$$D_s = \alpha D, \tag{3.2}$$

where $\alpha = 9.02 B_0 \cdot \frac{(1+0.45 Oh^{0.5})(1+0.4T^{0.7})}{(1+0.87 We^{1.67})^{0.6}}$ is assumed to be a weak function of $D$. Differentiating Equ. (3.1) with respect to time, we have the differential equation for $D(t)$,

$$\frac{dD}{dt} = -\frac{A^3 D}{3(N + A^3 n)} \frac{dn}{dt}. \tag{3.3}$$

From the energy conservation of parent droplets at time $t$, the work of external force $F$ on them converts into the surface energy and kinetic energy,

$$\frac{d}{dt}\left(\sigma_p n(t) \pi D_s^2 + \sigma_p N \pi D^2 + \frac{1}{2} M u_D^2\right) = N \cdot F \cdot u_r, \tag{3.4}$$

where $F = \rho_l \frac{1}{6} \pi D(t)^3 a \cdot u_r$. From Equ. (3.4), we have

$$\left(2\sigma_p \pi n A^2 D + 2\sigma_p \pi N D + \frac{N}{2} \pi \rho_l u_D^2 D^2\right)\frac{dD}{dt}$$
$$+ \sigma_p \pi A^2 D^2 \frac{dn}{dt} = \frac{1}{6} \pi N \cdot \rho_l D^3 a (u_r - u_D). \tag{3.5}$$

Combining Equs. (3.3) and (3.4), we have the following system of ordinary differential equations,

$$\begin{cases} \dfrac{dn}{dt} = f(n, D) \\ \dfrac{dD}{dt} = -\dfrac{A^3 D}{3(N + A^3 n)} f(n, D) \end{cases}, \tag{3.5}$$

where

$$f(n(t), D(t)) = \frac{\frac{1}{6} N \cdot \rho_l D a (u_r - u_D)}{\sigma_p A^2 - \frac{A^3}{3(N + A^3 n)}\left(2\sigma_p n A^2 + 2\sigma_p N + \frac{N}{2} \rho_l u_D^2 D\right)}. \tag{3.6}$$

By numerically solving (3.5), we can update the $n(t)$ and $D(t)$ each time step and put the newly detached child droplets in a new parcel. The velocity of child droplets $\boldsymbol{u}_s$ can be determined by the linear momentum conservation.

$$N \cdot D^3(t)\boldsymbol{u} = N \cdot D^3(t+dt)\boldsymbol{u}_D + n(t+dt)\boldsymbol{u}_s. \qquad (3.7)$$

Compared with the ReitzKH-RT model, the present model is free of the KH breakup time parameter $\tau_{KH}$ by introducing the ODE system, while $\tau_{KH}$ can be derived from the present model if we integrate $t$ from 0 to $\tau_{KH}$ in the breakup process, where the droplet diameter decreases from $D_0$ to $D_s$ as

$$\int_0^{\tau_{KH}} dt = -\int_{D_0}^{D_s} \frac{dD}{f(n,D)} \frac{3(N+A^3 n)}{A^3 D}, \qquad (3.8)$$

from which we have

$$\tau_{KH} = -\int_{D_0}^{D_s} \frac{3(N+A^3 n)}{f(n,D)A^3} \frac{dD}{D}. \qquad (3.9)$$

In the ReitzKH-RT model, $\tau_{KH} = \frac{3.726 B_1 D}{2\Omega_{KH}\Lambda_{KH}}$ involves a fitting parameter $B_1$ to align with experiments, and $B_1$ ranges from 1.71 to 100. The present model is able to degenerate to the ReitzKH-RT model when the Weber number is relatively low.

### 3.2. One-dimensional Detonation Velocity

To study the effects of droplet diameter and breakup models on one-dimensional two-phase detonation of n-heptane, we conducted simulations for four breakup models (TAB, PilchErdman, ReitzKH-RT, and present model) for droplet diameters of 30, 50, and 70 μm. To illustrate the reliability of the numerical method, we compare the basic parameters of the detonation wave at a droplet diameter of 30 μm with the experiments of Benmahammed [34], as shown in Table 3. The results show that, regardless of the breakup model used, the error is within 8%, indicating that the present computational methodology is applicable to the n-heptane two-phase detonation.

| Model | $U_D$ [m/s] | $U_D$ Errors | $P_{cj}$ [atm] | $P_{cj}$ Errors | $T_{cj}$ | $T_{cj}$ Errors |
|---|---|---|---|---|---|---|
| TAB | 1805 | 0.45 % | 17.50 | 7.41 % | 2795 | 1.27 % |
| PilchErdman | 1868 | 3.90 % | 18.30 | 3.17 % | 2828 | 0.11 % |
| ReitzKH-RT | 1833 | 2.00 % | 18.03 | 4.60 % | 2840 | 0.32 % |
| Present | 1815 | 1.00 % | 17.77 | 5.98 % | 2827 | 0.14 % |
| No Breakup | 1805 | 0.45 % | 17.65 | 6.61 % | 2813 | 0.64 % |
| Experiment[33] | 1797 | - | 18.90 | - | 2831 | - |

Table 3 Comparison of thermodynamic parameters of one-dimensional detonation wave with experiment [33].

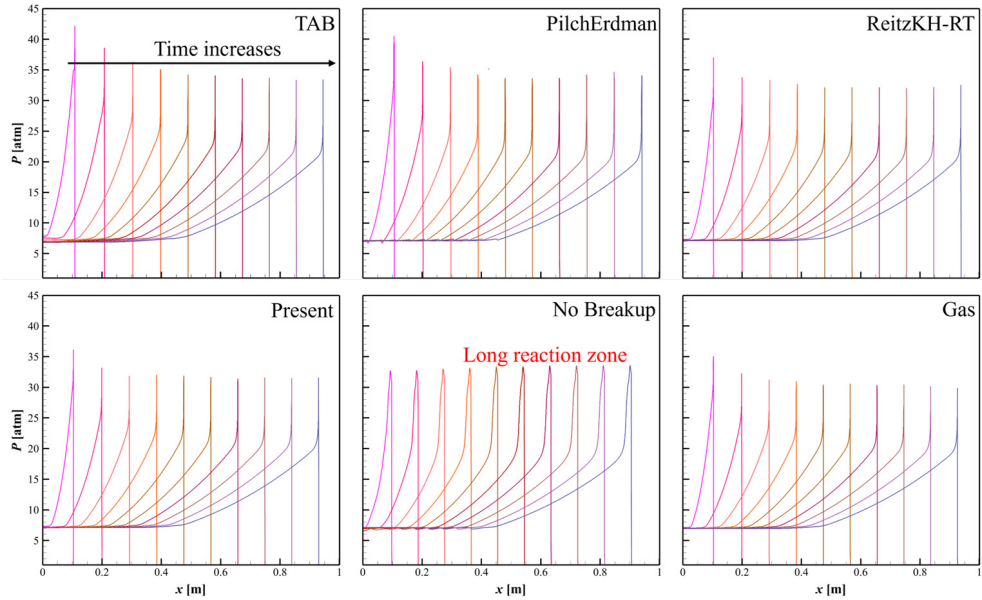

Fig.5 Evolution of pressure profiles of one-dimensional detonation wave for $D_0$= 30 μm.

Fig.5 illustrates the pressure variation across the computational domain at various time intervals for the different cases, all featuring n-heptane droplets with $D_0 = 30$ μm. The average thermodynamic parameters of the detonation waves are given in Table 3. It is seen that that the detonation waves of the different cases reach a steady state after $x = 0.4$ m and propagate self-sustainably. Without a droplet breakup model, the simulated pressure peak of the shock wave is significantly broader, with a longer delay from droplet breakup to chemical reaction. However, with a droplet breakup model, the pressure peak area is narrower owing to the narrower regions of leading shock wave, droplet breakup, evaporation, and reaction. In addition, the pressure curves of detonation waves for different breakup models are similar to each other and to those of the gas-phase n-heptane simulation. As shown in Table 3, different macroscopic thermodynamic quantities related to the characteristics of detonation waves are consistent regardless of without/with droplet breakup models and of different models.

|  | Model | $U_D$ [m/s] | $P_{cj}$ [atm] | $T_{cj}$ [K] |
|---|---|---|---|---|
| $D_0 = 50$ μm | TAB | 1800 | 17.41 | 2787.34 |
|  | PilchErdman | 1804 | 17.46 | 2803.64 |
|  | ReitzKH-RT | 1834 | 18.04 | 2840 |
|  | Present | 1808 | 17.63 | 2810.64 |
| $D_0 = 70$ μm | TAB | 1800 | 17.56 | 2784.16 |
|  | PilchErdman | 1797 | 17.41 | 2793.33 |
|  | ReitzKH-RT | 1841 | 18.10 | 2841.69 |
|  | Present | 1802 | 17.50 | 2796.48 |

Table.4 Comparison of thermodynamic parameters of one-dimensional detonation waves for the different breakup models at $D_0$ = 50 and 70 μm

To further illustrate the results, Table 4 presents the thermodynamic quantities of the different cases under the conditions of n-heptane droplets of 50 and 70 μm. Under such conditions, the detonation waves without breakup models are not self-sustainable. The propagation speed of detonation waves simulated by using the TAB and ReitzKH-RT models does not decrease as the droplet diameter increases, while it slightly decreases by using the PilchErdman and Present models. The detonation wave propagation velocities do not show significant changes among different breakup models.

The detonation wave relation can be used to explain the similarity of macroscopic thermodynamic quantities in the different cases. For the gas-phase detonation wave at CJ (Chapman and Jouguet) state [35], we use the subscripts 0 and 1 to represent the pre-wave and post-wave gas states respectively. With the assumptions of $p_0 \ll p_1$ and the constant specific heats of reactants and products, the CJ detonation speed can be expressed as

$$U_{CJ} = \sqrt{2(\gamma_1^2 - 1)c_{p1}\left(\frac{q}{c_{p1}} + \frac{c_{p0}}{c_{p1}}T_0\right)}. \qquad (3.10)$$

It is seen that the CJ detonation speed is a thermodynamic quantity only related to the thermodynamic parameters before and after the detonation wave. In addition, the detonation propagation velocity $U_D$ is close to $U_{CJ}$. When it comes to the two-phase detonation, Fig.6 gives the curves of temperature and $H_2O$ mass fraction near the detonation wave front for the different breakup models. Comparing the curves of different breakup models with the same initial droplet diameter, we can see that there is a stepwise increase in temperature for the TAB and PilchErdman models. This result implies that there is a certain induction zone after the shock wave, whereas the induction zone is not apparent for the other two models, and that the combustion couples with the detonation wave more closely. As for the TAB and PilchErdman breakup models, the larger the diameter of the droplet, the longer the corresponding induced zone. However, the temperature and mass fractions of the product $H_2O$ are very close for all the models after the detonation wave, which indicates that n-heptane fuel is fully reacted in these cases. Therefore, the total amount of heat released from the chemical reaction is almost the same, and the temperature is accordingly very close. In combination with the expression for the CJ detonation wave velocity, the similar temperature explains why the different cases here have similar detonation velocities.

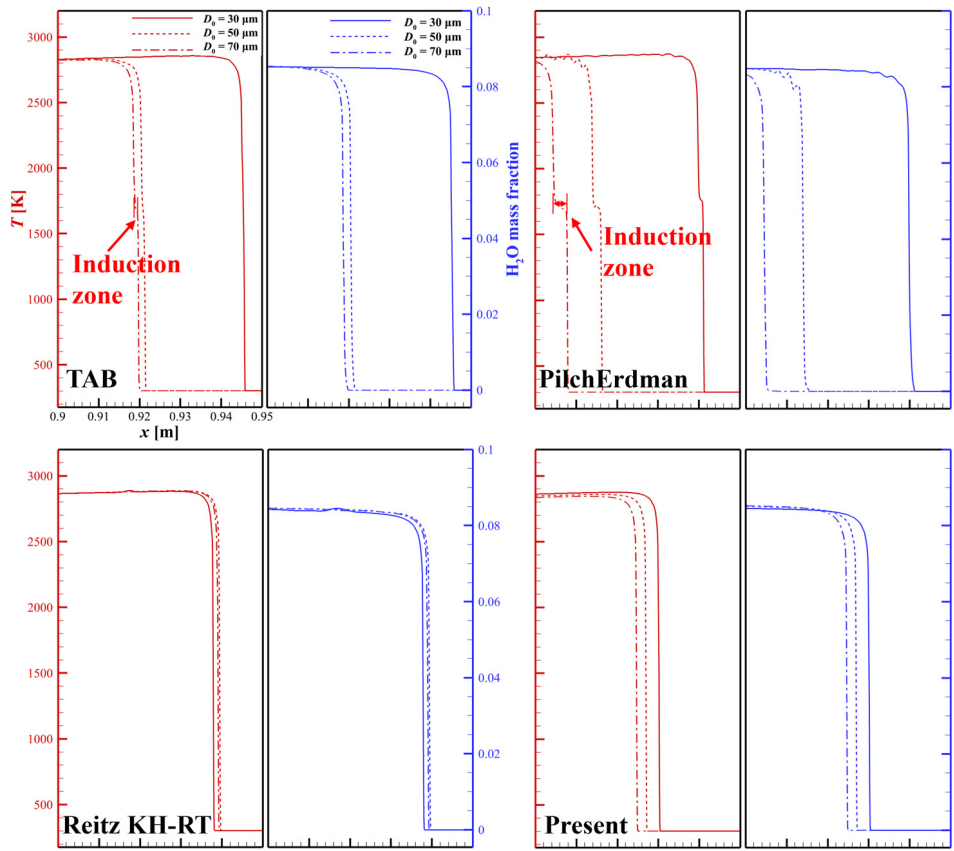

Fig.6 Temperature and $H_2O$ mass fraction curves near the detonation wavefront for the different breakup models

## 3.3. *Two-dimensional Detonation Cell Size*

This section considers the two-dimensional detonation of different breakup models under different droplet diameters. Similar to the one-dimensional analysis, this section also considers a combination of three droplet diameters and four breakup models. It is worth noting that for the ReitzKH-RT model, the simulations are carried out for two $B_1$ parameters: the commonly used value $B_1 = 40$ and the recommended value $B_1 = 1.71$ in the reference [10, 36].

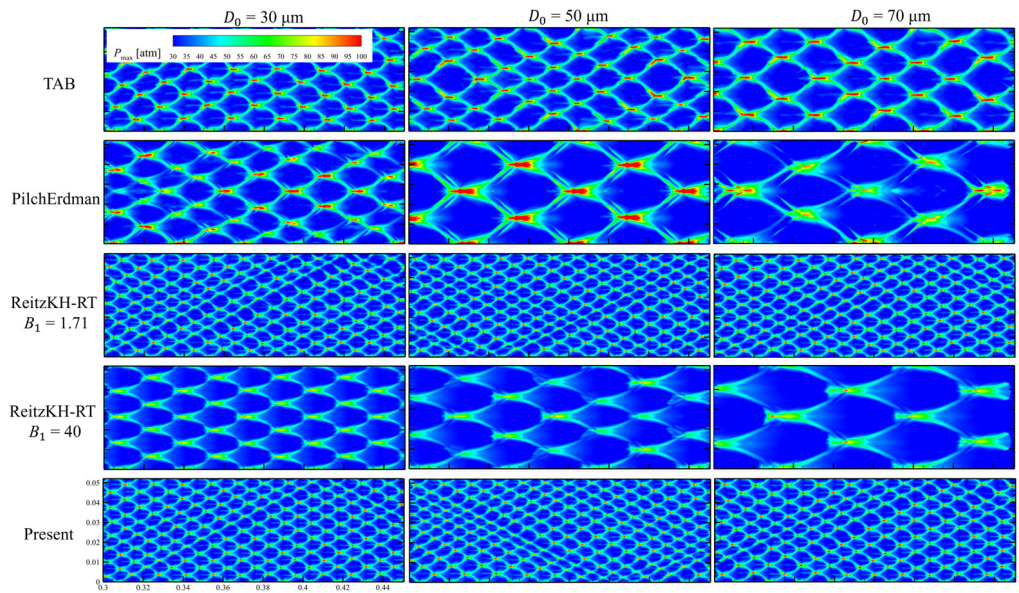

Fig.7 Two-dimensional detonation wave cell structure for the different initial droplet diameters and breakup models.

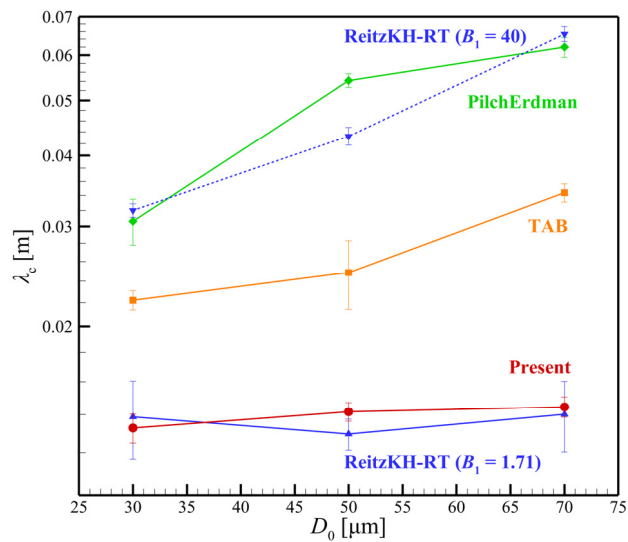

Fig.8 Average sizes of two-dimensional detonation wave cell for the different initial droplet diameters and breakup models.

Fig.7 shows the detonation cell structure contour for each simulation case. The vertical comparison reflects the influence of different breakup models on the detonation cell size for the same droplet diameter. To quantify the comparison results, Fig.8 presents the average detonation cell size for different cases. Specifically, for the same initial droplet diameter, different breakup models give significantly different detonation cell sizes. The PilchErdman model and ReitzKH-RT model with $B_1 = 40$ generate larger detonation cell sizes, but the present model and the ReitzKHRT model with $B_1 = 1.71$ generated smaller detonation cell sizes, and the detonation cell sizes of the TAB model are in between. Moreover, the detonation cell sizes generated by the ReitzKH-RT model are sensitive to the adjustable parameter $B_1$, which controls the KH breakup time. The ReitzKH-RT model with $B_1 = 40$ is similar to the Pilch-Erdman model, and the ReitzKH-RT model with $B_1 = 1.71$ is similar to the present model. The present model, generate the results similar to those by the ReitzKH-RT model at recommended values without introducing the KH breakup time adjustable parameter $B_1$, instead by using an ODE system based on conservation laws to model the breakup process. This demonstrates the applicability and generality of the present model. The horizontal comparison results show that the impact of droplet diameter on different models varies. For the PilchErdman, the TAB, and the ReitzKH-RT models with $B_1 = 40$, detonation cell size increases significantly with increasing initial droplet diameter. For the ReitzKH-RT model with $B_1 = 1.71$, the detonation cell sizes have little difference with increasing droplet diameter. For the present model, the detonation cell sizes increase slightly with the initial droplet diameter.

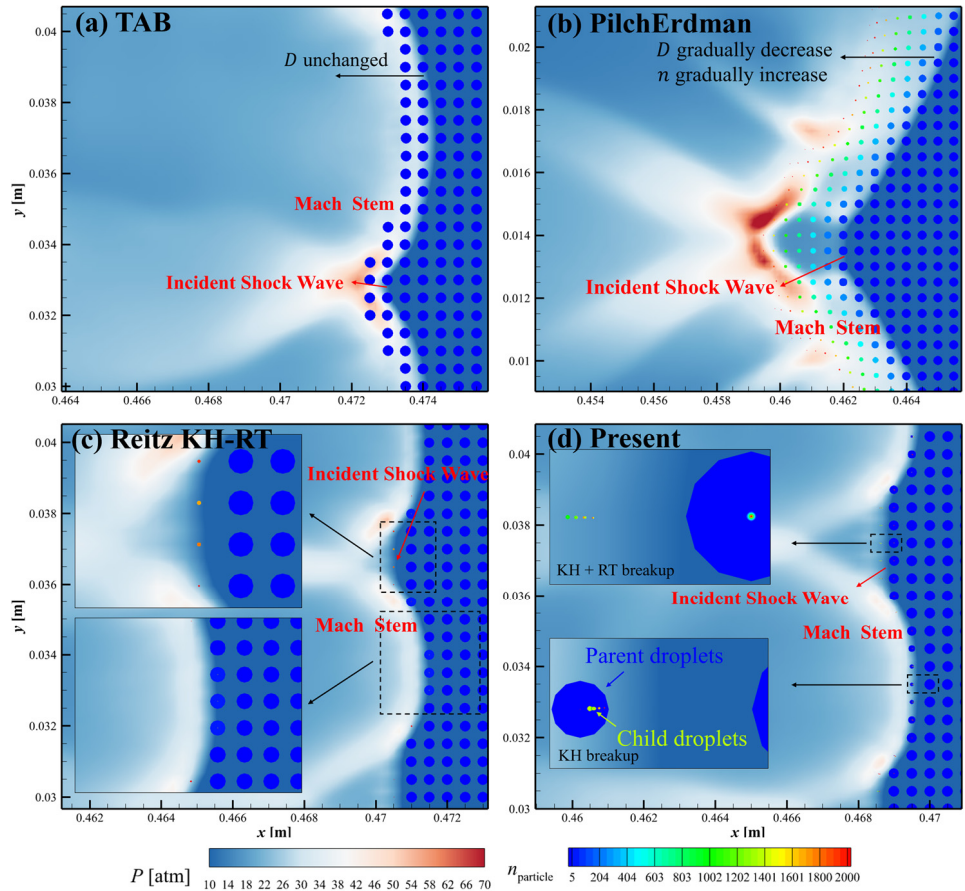

Fig.9 Contours of pressure and droplet size distribution for the different breakup models at $D_0 = 50$ μm (a) TAB model (b) PilchErdman model (c) Reitz KH-RT model (d) Present model

To analyze the above observations, the droplet distributions generated by different breakup models near the detonation wave are shown in Fig. 9, where the pressure contour and droplet distribution are at $t = 0.25$ ms, when the detonation wave propagates in a self-sustaining and stable manner. The relative size of the droplets represents the droplet diameter in different droplet parcels, while the color represents the number of droplets in the same droplet parcel. It should be noted that the droplets shown in the figure are enlarged proportionally for clarity. For the TAB model, the droplet diameter and number in the droplet packet remain unchanged for a certain period after passing through the shock wave and the droplets are in the stage of deformation, which is consistent with the physical image depicted by the TAB model. The droplets instantaneously break into several extremely small droplets, and they rapidly evaporate to participate in the gas-phase reactions, so no small-diameter droplets are observed. For the PilchErdman model, the longer the cumulative time of droplets passing through the shock wave, the smaller the droplet diameter and the more droplets there are, which reflects the 'chain breaking process' described by this model. For the ReitzKH-RT model, the droplet parcels have small diameters and large particle numbers after the incident shock wave. The breakup process is faster after the Mach stem and small droplets evaporate quickly without being visible in the figure. For the present model, both large droplets and small droplets peeled off by the KHI wave can be observed after the incident shock wave, indicating that both the KH and RT breakup have occurred here. Furthermore, the coexistence of parent and daughter droplets is observed after the Mach stem, indicating that the KH breakup is the dominant mechanism at this stage.

To consider the impact of the different droplet distribution due to the different breakup models on the chemical reactions, we present in Fig.10 the heat release rate and droplet distribution contour. It can be observed that the distance between the chemical reaction heat release zone and the wave head $l_i$, namely the induction zone, varies for the different breakup models, which is the main factor affecting the detonation cell sizes. For the TAB model, the heat release zone is concentrated at where large droplets disappear and instantaneously break into small droplets with the same size as the minimum diameter droplet given by this model. For the other models, the heat release zone is concentrated in the region where the smallest droplets appear. Since the difference between the different cases lies only in the breakup model, $l_i$ actually represents the breakup length of each model and is estimated as

$$l_i \approx u_p \tau_b, \tag{3.11}$$

The breakup time, $\tau_b$, is also the timescale for the generation of the smallest droplets, which generally have very quick evaporation and combustion. Consequently, the droplet breakup process, as the rate-controlling process, is reflected in the different spatial scales of $l_i$ in the contour. From Fig.10 and Fig.8, we have

$$\begin{aligned} l_{i,ReitzKH-RT\ (B_1=40)} &\approx l_{i,PilchErdman} > l_{i,TAB} \\ &> l_{i,Present} \approx l_{i,ReitzKH-RT\ (B_1=1.71)}, \end{aligned} \tag{3.12}$$

and

$$\begin{aligned} \lambda_{c,ReitzKH-RT\ (B_1=40)} &\approx \lambda_{c,PilchErdman} > \lambda_{c,TAB} \\ &> \lambda_{c,Present} \approx \lambda_{c,ReitzKH-RT\ (B_1=1.71)}, \end{aligned} \tag{3.13}$$

which show that $\lambda_c$ is positively correlated with $l_i$ and hence $\tau_b$. For the PilchErdman and the ReitzKH-RT model with $B_1 = 40$, the longest timescale is required for the generation of the smallest droplets. Therefore, the total length of the chemical reaction zone behind the detonation wave head is longer, resulting in larger detonation cell sizes. For the ReitzKH-RT model with $B_1 = 1.71$, the timescale for the KH breakup is set to be short, leading to the rapid generation of small droplets and smaller cellular structures. This result also indicates that the ReitzKH-RT model is sensitive to $B_1$. As for the present model, both large and small droplets coexist. Since the small droplets start to form, evaporate, and react immediately after the leading shock wave because of the KH breakup process, the corresponding chemical reaction zone is short, resulting in smaller detonation cell sizes. The TAB model falls between the other models, with intermediate detonation cell sizes. These results reveal the reasons for the different detonation cell sizes in the different breakup models, and further experiments are needed to determine which physical process is closer to the reality.

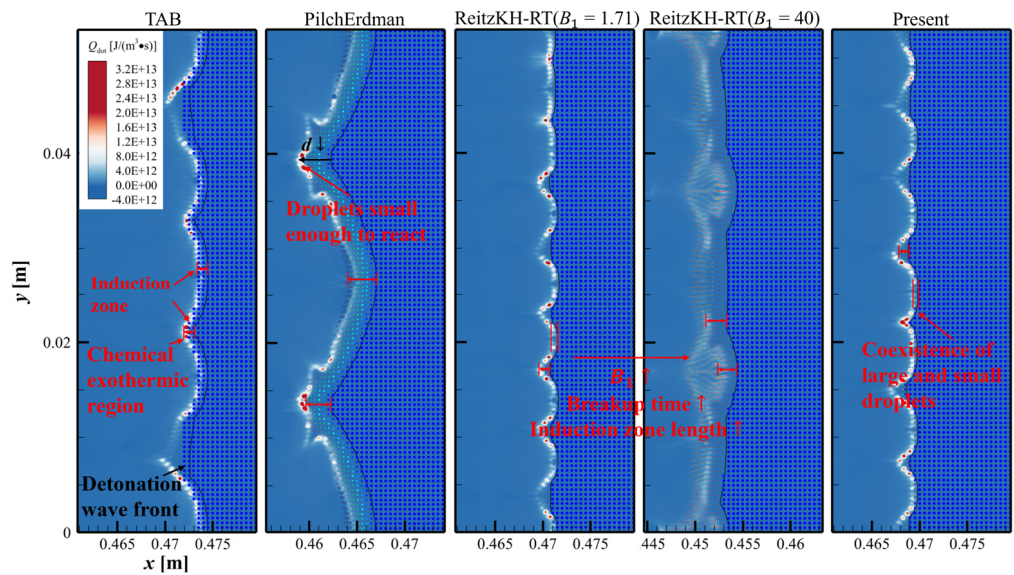

Fig.10 Contours of chemical reaction heat release rates and droplet distribution for the different breakup models at $D_0 = 50$ μm

## 4. Conclusions

The liquid-fueled detonation is a supersonic flow that involves complex gas-liquid interactions and chemical reactions. The timely breakup and vaporization of liquid droplets are crucial to sustain the detonation. In this work, we proposed a dynamic droplet breakup model based on the conservation laws and implemented it in our open-source two-phase supersonic flow solver based on rhoCentralFoam in OpenFOAM. Then we compared this model to three existing models for different droplet initial diameters in the one- and two-dimensional detonation problems.

The simulation results show that the present model works well in predicting detonation parameters in one-dimensional detonation. In two-dimensional detonation, the present model produces similar detonation cell sizes compared with the ReitzKH-RT model for the parameter suggested in the literature, eliminating the need for introducing and adjusting the KH breakup time parameter.

For one-dimensional detonation, the different droplet breakup models mainly affect the length of the chemical reaction-induced zone but have a slight effect on the propagation velocity of the detonation wave. Because the detonation velocity is mainly determined by the thermodynamic properties before and after the detonation wave, it is not substantially affected as long as the detonation can be sustained in the simulations with the different breakup models. For two-dimensional detonation, the different droplet diameters and breakup models significantly influence the cell size of detonation waves. Larger initial droplet diameters result in larger cell sizes. Moreover, distinct breakup models yield different physical processes of droplet breakup, leading to different distributions of droplet diameter, evaporation rates, and heat release rates. The chemical reaction zone is concentrated in the region where the smallest droplets appear, and the spatial scale of the formation of small droplets after the detonation wave determines the scale of the chemical reaction zone, which is reflected in the detonation cell size. So a longer droplet breakup length will result in a larger detonation cell size.

For future work, we note that the existing breakup models do not describe the same physical process and are suitable for different ranges of flow parameters, a unified model is always desirable, and further experiments are needed for model validation. In many two-phase detonation problems, the distribution of droplets is usually polydisperse, and the droplets of different sizes may have nonuniform motion and therefore have chance to collide with each other, which necessitates further investigation of droplet collision models.

**ACKNOWLEDGMENTS**

This work was supported by the National Natural Science Foundation of China (Grant No. 52176134). The work at the City University of Hong Kong was additionally supported by grants from the Research Grants Council of the Hong Kong Special Administrative Region, China (Project No. CityU 15222421 and CityU 15218820).